\documentclass[10pt]{article}
\usepackage{abstract}
\usepackage{graphicx}
\usepackage{xcolor}
\usepackage{caption}
\usepackage{mathtools}
\usepackage{tikz}	
\usepackage{amsmath, amsthm, amssymb}
\usepackage[english]{babel}
\usepackage{geometry}
\usepackage{floatrow}
\usepackage{subcaption}
\usepackage{booktabs}
\usepackage{braket}
\usepackage{authblk}
\usepackage[thinlines]{easytable}
\usepackage[utf8]{inputenc} 
\usepackage{ulem}
\graphicspath{{figures/}}
\newcommand{\HH}{{\mathcal{H}}}

\newcommand{\UU}{\textrm{U}}
\newcommand{\rf}{\textrm{rf}}
\newcommand{\dd}{\textrm{d}}

\newcommand{\rr}{\textrm{r}}

\newcommand{\eff}{\textrm{eff}}
\newcommand{\II}{\textrm{I}}

\newcommand{\Fou}{\mathcal{F}}
\newcommand{\iFou}{\mathcal{F}^{-1}}

\newcommand{\CS}{\textrm{CS}}

\newlength{\arrow}
\providecommand{\keywords}[1]{\textbf{\textit{Keywords:}} #1}

\begin{document}
\bibliographystyle{unsrt}
\title{Interaction frames in solid-state NMR: A case study for chemical-shift-selective irradiation schemes}

\author[1]{Matías Chávez}
\affil{\small Physical Chemistry, ETH Zürich, Vladimir-Prelog-Weg 2, 8093 Zürich, Switzerland}

\author[1]{Matthias Ernst \thanks{Corresponding author (maer@ethz.ch)}}%

\maketitle
\begin{abstract}
Interaction frames play an important role in describing and
understanding experimental schemes in magnetic resonance. They are
often used to eliminate dominating parts of the spin Hamiltonian,
e.g., the Zeeman Hamiltonian in the usual (Zeeman) rotating frame, or the radio-frequency-field (rf) Hamiltonian to describe the efficiency of decoupling or recoupling sequences. Going into an interaction frame can also make parts of a time-dependent Hamiltonian time independent like the rf-field Hamiltonian in the usual (Zeeman) rotating frame.  Eliminating the dominant term often allows a better understanding of the details of the spin dynamics. Going into an interaction frame can also reduce the energy-level splitting in the Hamiltonian leading to a faster convergence of perturbation expansions, average Hamiltonian, or Floquet theory. Often, there is no obvious choice of the interaction frame to use but some can be more convenient than others. Using the example of frequency-selective dipolar recoupling, we discuss the differences, advantages, and disadvantages of different choices of interaction frames. They always include the complete radio-frequency Hamiltonian but can also contain the chemical shifts of the spins and may or may not contain the effective fields over one cycle of the pulse sequence.
\end{abstract}

%
%
\section{Introduction}
Interaction-frame transformations are an essential tool and frequently
utilized in magnetic resonance \cite{Abragam1961, Ernst:1990vd, Haeberlen:1976uz, Mehring:1983wm}.
Almost all Hamiltonians used in high-field NMR are  expressed in an interaction
frame generated by a rotation about the Zeeman Hamiltonian, an
interaction frame that is often called the rotating frame
\cite{Abragam1961, Ernst:1990vd, Haeberlen:1976uz, Mehring:1983wm}. The main advantage of the technique is the
simplification of the spin dynamics of the system described by the
Liouville-von Neumann equation \cite{Neumann:1932}, without introducing
approximations. This is often necessary to obtain a solution with a
feasible amount of computational effort or to clarify the underlying
physical processes. Different interaction frames can be used to
highlight different aspects of the system and can provide advantages
for subsequent treatments like operator-based perturbation theory \cite{Vleck1929,Primas:1961vi,Primas:1963tg}, average Hamiltonian theory \cite{Haeberlen:1968wu, Haeberlen1976} or Floquet theory
\cite{Floquet:1883vq, Shirley:1965tn, Vinogradov:2004es, Scholz:2010hq, Ivanov2021} by reducing the energy-level spacing of the system. \\
The two most prominent analytical approaches in magnetic resonance, (standard) Floquet theory and average Hamiltonian theory (AHT)
are restricted to periodic interaction frames that are repeated
indefinitely, i.e., interaction frames following a periodic motion. We have recently
introduced a continuous Floquet approach \cite{Chavez:2022} that
is also compatible with an arbitrary interaction-frame transformation
and sequences, that can be limited in duration enabling, therefore, new possibilities and raising again the question of the choice of the interaction frame.
                                                                                                                                                                                                                                                                                                                                                                                                                                                                                                                                                                                                                                                                                                                                                                                                                                                                                                                                           \\
In this paper we analyze different interaction-frame transformations in the context of radio-frequency (rf) irradiation and show how they can be used to answer different questions.  To illustrate the techniques, we use the example of chemical-shift selective irradiation schemes in solid-state NMR under magic angle spinning (MAS) \cite{Agarwal:2018kc, Potnuru:2020ez, Duong:2020gc, Zhang:2020fq, Potnuru:2021cn, Xiao:2021ih, Nimerovsky:2022bs}. Different interaction frames emphasize different aspects of a problem and effects will be observed in different orders of the series expansion. Often such different viewpoints enable a more complete understanding of the spin dynamics. In general, excluding an interaction from the interaction-frame transformation, hence isolating it, can be used to analyze its effects on the dynamics of the system. For example, the isotropic chemical shift can be excluded from the interaction frame and, therefore, the effect of the chemical shift is directly visible in this frame making the analysis simpler. This approach is explored in the context of homonuclear recoupling in section \ref{sec:chemical-shift-frame}.
                                                                                                  \\

                                                                                                  In a slightly different but related approach, the chemical shift is included in the interaction-frame transformation but the effective fields over one period of the irradiation scheme are taken out of the interaction frame by a second transformation (see section \ref{sec:effective-frame}). Such a frame was first described in \cite{Basse:2016ir} and later used for the description of homonuclear recoupling under MAS \cite{Shankar:2017ck} for the description of TOCSY transfer efficiencies \cite{Nielsen:2019kf}. Although isolating terms of the Hamiltonian from the interaction frame gives physical insights how the sequence works, the Hamiltonian becomes simpler as more terms are included into the interaction frame. A simple Hamiltonian is more convenient for subsequent treatments like perturbation theory, especially in higher orders. In addition, the accuracy of a specific order of the perturbation theory will increase when more of the dynamic is included into the interaction frame. Therefore, the simplest Hamiltonian, leads to the most precise and convenient perturbation treatment, which is the interaction frame including the complete rf-field and chemical-shift Hamiltonian.                                                                                       
%
%
\section{Theory}
\subsection{General formalism}
Interaction frames are unitary time-dependent transformations mediated by a propagator often based on a subset of the terms making up the total Hamiltonian.
In general, a time-dependent Hamiltonian transforms under a unitary propagator as
\begin{align}\label{eq:general_transformation}
    {\HH'}(t) = \UU^\dagger(t)\HH(t)\UU(t) - i\, \UU^\dagger(t)\Dot{\UU}(t)
\end{align}
Here, the interaction-frame Hamiltonian is characterized by an apostrophe. In the following we have basically only interaction-frame Hamiltonians, therefore we drop the apostrophe from now on for easier readability.
The second term of the equation is sometimes referred as the Coriolis term and is only dependent on the propagator $\UU(t)$. It can also be seen as a correction term in the Liouville-von Neumann equation when going into an interaction frame. This term is responsible for the elimination of the terms of the Hamiltonian that are included in the interaction-frame transformation. However, if the transformation is not generated by the exponential of a term that is part of the Hamiltonian, the Coriolis term introduces an additional term in the Hamiltonian as in section \ref{sec:effective-frame}.
We will restrict our discussion here to interaction-frame transformations based on one-spin operators, i.e., radio-frequency irradiation and chemical-shift terms. The rotations of the interaction frame compared to the initial frame, i.e. the frame at time $t = t_0$ can be encoded in the matrix of one-spin coefficients $a_{\mu\nu}(t)$ called interaction-frame trajectory, such that 
\begin{align}
\UU^\dagger(t)\,\II_\mu(t_0)\,\UU(t)
= \sum\limits_{ \chi}a_{\mu\chi}(t)\,\II_\chi(t_0)
\end{align}
where $\II_\mu(t_0)$ indicate the spin operators in the initial frame with $\mu,\chi \in \{x,y,z\}$ or $\mu,\chi \in \{+,-,z\}$ or any other complete orthonormal basis set. 
Using continuous Floquet theory \cite{Chavez:2022}, the Hamiltonian in the interaction-frame is 
expressed as an (element-wise) inverse Fourier transform, whereas the spatial modulations due to MAS are encoded with a Fourier series expansion.
We define the Fourier transform as
\begin{align}
  \label{eq:FourierTransform}
&\HH(t) = \iFou\{{\widehat{\HH}(\Omega)}\} =  \tau \int\limits_{-\infty}^\infty\widehat{\HH}(\Omega)e^{i \Omega t}\dd\Omega\\
&\widehat{\HH}(\Omega) = \Fou\{{\HH(t)}\} = \frac{1}{2\,\pi\, \tau} \int\limits_{0}^\tau\HH(t)e^{-i \Omega t}\dd t
\end{align} 
where $\tau$ is the duration of the rf-field irradiation \cite{Chavez:2022}.
As a result the interaction-frame Hamiltonian takes the general form
\begin{align}
{\HH}(t) = \tau \sum\limits_{n = -2}^2 \, \int\limits_{-\infty}^\infty \widehat{\HH}^{(n)}(\Omega)\, e^{i n\omega_{\rr} t}
  e^{i \Omega t}\, \dd \Omega
\end{align}
where $\HH^{(n)}(\Omega)$ is the frequency-domain Hamiltonian, i.e. the Fourier transform of the Hamiltonian.
Here the superscript $(n)$ indicates the $n\,$th Fourier coefficient resulting from the modulation due to MAS rotation with the frequency $\omega_\textrm{r}$. In the case of MAS of spin-1/2 nuclei $n\in\{-2,-1,0,1,2\}$, since the only spatial dependence stems from the dipolar coupling or the chemical-shielding tensor which are rank-2 space tensors.
The Fourier transformation, indicated by the a wide hat, is in respect to the spin-space modulation due to the interaction-frame transformation.
Notice, that the spin-space modulations can encode multiples of the modulation frequency as well as the modulations by the effective fields.
Under van Vleck perturbation theory  \cite{Vleck1929,Primas:1961vi,Primas:1963tg}, the first-order effective Hamiltonian has the form 
\begin{align}
\bar{\HH}^{(1)} = \sum^2_{n=-2} \widehat{\HH}^{(n)}(n\omega_\textrm{r})
\end{align}
which is the frequency-domain Hamiltonian summed at the resonance conditions $\Omega = n \omega_\textrm{r}$.
Based on this general formalism we consider a simple homonuclear spin system including only isotropic chemical shifts and dipolar couplings under general rf-field irradiation and MAS. The MAS time-dependent rotating-frame Hamiltonian is given by
\begin{align}\label{eq:rotating-frame-Hamiltonian}
\HH(t) \notag
& =
\sum_{i,j}\omega_{ij}(t)\,(3 \, \textrm{I}_{iz}\textrm{I}_{jz}-\vec{\textrm{I}}_i \vec{\textrm{I}}_j) \notag
\\
& \quad +
     \sum\limits_{i=1}^{N} \omega_{i}\textrm{I}_{iz}
\\
& \quad +
\sum\limits_{i=1}^{N} \omega_1(t)\big(\cos(\phi(t))\II_{ix}+\sin(\phi(t))\II_{iy}\big)  \notag
\end{align}
Here we  have already represented the MAS-modulated, time-dependent dipolar coupling as a Fourier series by
\begin{align}
    \omega_{ij}(t) =  \sum\limits_{n = -2}^2 \omega_{ij}^{(n)}e^{i n \omega_{\rr} t}
    \label{eq:dipolar-coupling}
\end{align} 
%
%
\subsection{Interaction frames}
\begin{figure}[H]
	\caption{Schematic illustration of the three different choices of the interaction frames. In (IF1), the Hamiltonian is transformed into the interaction frame with the radio-frequency Hamiltonian. After Fourier transformation and using first-order Floquet theory, we end up with a scaled chemical-shift term and a scaled  coupling term which consists only of second-rank contributions. In (IF2), the Hamiltonian is transformed into an interaction frame using the chemical-shift and the dipolar-coupling Hamiltonian. After Fourier transformation and using first-order Floquet theory, we obtain a scaled coupling term that contains zeroth-, first- and second-rank tensor components. The (IF3) case uses the same interaction-frame transformation as (IF2). However in (IF3) we apply a second transformation to eliminate the effective fields from the interaction frame thus reintroducing them in the Hamiltonian. After Fourier transformation and first-order Floquet treatment, we obtain an effective Hamiltonian that contains the time-independent effective fields and the scaled coupling term that can again contain zeroth-, first- and second-rank tensor components.} 
	\label{fig:frames}
	{\includegraphics[width=17cm]{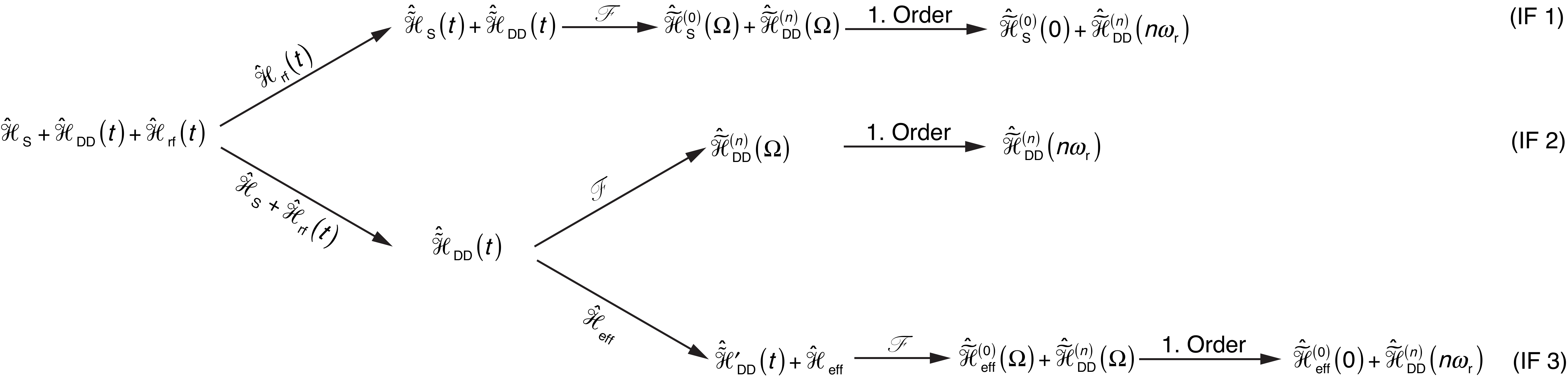}}
\end{figure}

At this stage we have to decide which interaction frame we use for the analysis, i.e., which parts of the Hamiltonian to included into the interaction-frame transformation. There are two obvious choices: (i) We can either use only the rf-field Hamiltonian (see Eq. \eqref{eq:rotating-frame-Hamiltonian}) or (ii) we can use the isotropic chemical-shift term and the rf-field Hamiltonian (see Eq. \eqref{eq:rotating-frame-Hamiltonian}). This is illustrated in Fig. \ref{fig:frames} where three different interaction frames are shown, one based on the transformation by the rf-field Hamiltonian only (IF1) and two based on the combined transformation by the rf-field and the chemical-shift Hamiltonian (IF2 and IF3).
In the first case, we obtain an interaction-frame Hamiltonian that contains the modulated dipolar coupling and the chemical-shift terms. After Fourier transformation and Floquet theory we obtain a first-order effective Hamiltonian that contains the scaled chemical shifts and the resonant parts of the dipolar coupling. In the second case we only obtain a modulated generalized coupling term that contains zeroth-, first-, and second-rank two-spin terms.
This is due to the fact, that the two spins are transformed by different interaction-frame trajectories due to the different chemical shifts. We can either directly apply Fourier transformation and Floquet theory on this modulated coupling and obtain a first-order effective Hamiltonian that contains all possible resonant two-spin terms (IF2).
Alternatively we can take the effective fields out of the interaction frame (IF3) by a second interaction-frame transformation and obtain after Fourier transformation and Floquet theory a first-order effective Hamiltonian that contains the two effective fields and the resonant effective coupling terms. We will discuss these three cases in the next three sub sections.
%
%
\subsubsection{Interaction frame without chemical shift}
\label{sec:chemical-shift-frame}
In the first approach the isotropic chemical shift is not included in the interaction-frame transformation, to facilitate a better understanding of its effect on the transfer efficiency.
The transformations of the Hamiltonian can be seen in Fig. \ref{fig:frames} part (IF1). 
In the first step we go into an interaction frame with the rf-field Hamiltonian only, and leave the homonuclear coupling and the chemical-shift terms in the Hamiltonian.
The interaction-frame transformation is accomplished by the unitary operator
\begin{align}
    \UU(t) = \hat{T} \exp\left(-i \int\limits_{t_0}^{t} \HH_\textrm{rf}(t')\, d t'\right)
\end{align}
leading to the interaction-frame Hamiltonian 
\begin{align}
{\HH}(t)
=
\sum_{i,j}
 \omega_{ij}(t)\sum_{\mu,\nu}
\,{A_{\mu \nu}(t)\, \II_{i\mu} \II_{j\nu}}
+ \sum_{i=1}^N\sum_{\chi}\omega_{i} \, a_{z\chi}(t) \, \II_{i\chi}
\end{align}
The coupling coefficients $A_{\mu\nu}(t)$ are defined as
\begin{align}\label{eq:coupling-coefficient}
A_{\mu\nu}(t) = 2\,a_{z\mu}(t) a_{z\nu}(t) - \frac{1}{2}\left(a_{+\mu}(t) a_{-\nu}(t)+a_{-\mu}(t) a_{+\nu}(t)\right)
\end{align}
At this stage we can calculate the first-order effective Hamiltonian using Fourier transformation and applying van Vleck perturbation theory, resulting in
\begin{align}\label{eq:HamiltonianCS}
\bar{\HH}^{(1)} = \sum_n {\HH}^{(n)}(n\omega_\textrm{r})
=
\sum_n
\sum_{i,j}
\omega^{(n)}_{ij}\sum_{\mu,\nu}\widehat{A}_{\mu \nu}(n\omega_\textrm{r})\, \II_{i\mu} \II_{j\nu}
+ \sum_{i=1}^N\sum_{\chi}\omega_{i} \widehat{a}_{z\chi}(0) \, \II_{i\chi}
\end{align}
Note, that the coefficients $\widehat{A}_{\mu \nu}(\Omega)$ and $\widehat{a}_{z\chi}(\Omega)$ are identical for all spins, since the chemical shift is not included in the interaction frame and, therefore, all spin share a single frame of reference.
It is convenient to align the effective chemical-shift vector $\vec a_{z}(0) = (\widehat{a}_{z,x}(0), \widehat{a}_{z,y}(0), \widehat{a}_{z,z}(0))^T$ and the z axis by a rotation. This will, however, lead to an additional mixing of the components  of $\widehat{A}_{\mu \nu}(\Omega)$.
%
%
\subsubsection{Complete interaction frame}\label{sec:complete-frame}
The second interaction frame we consider is the complete interaction frame where the chemical shift and the rf-field Hamiltonian are included into the interaction-frame transformation. As shown in Fig. \ref{fig:frames} after we go into the interaction frame we obtain only a modulated coupling term.
The interaction-transformation is mediated by the unitary operator 
\begin{align}
    \UU(t) = \hat{T}\exp\left(-i \int_{t_0}^t \left[ \HH_\rf(t') + \HH_{\CS}\, \right] \dd t'\right)
\end{align}
Notice that every spin with a different chemical shift has its own interaction frame, because the isotropic chemical shift is included in the transformation.  
We end up with only a modulated effective coupling term that can have contributions from zeroth-, first-, and second-rank tensor components.
\begin{align}\label{eq:Hamiltonian-full-frame}
{\HH}(t)= \sum_{i,j}
\omega_{ij}(t) \sum_{\mu,\nu}
A^{(ij)}_{\mu \nu}(t)\, \II_{i\mu} \II_{j\nu} 
\end{align}
In this case, the coupling coefficients $A^{(ij)}_{\mu \nu}(t)$ depend on the indices of the spins, since for different spins we will have different interaction-frame trajectories due to differences in the chemical shifts. Notice that $A^{(ij)}_{\mu \nu}(t)$ are  the weighting factor of the two-spin operators in the interaction frame and encode the information of the rf-field irradiation as well as the chemical shift.
To obtain an effective 
Hamiltonian, we transform to the frequency domain and
apply van Vleck perturbation theory to first order resulting in
\begin{align}
\bar{\HH}^{(1)}
               & = \sum_{i,j} \sum_n
\omega^{(n)}_{ij} \sum_{\mu,\nu}
\widehat{A}^{(ij)}_{\mu \nu}(n \omega_{r})\, \II_{i\mu} \II_{j\nu}
\end{align}
In this case, the effective Hamiltonian contains only a coupling term and the magnitude of the polarization transfer will only depend on the relative and absolute magnitude of the different two-spin terms. This is in contrast to IF1 (\textit{vide supra}) and IF3 (\textit{vide infra}) where we also have a scaled shift term that can truncate the polarization transfer mediated through the coupling term.
%
%
\subsubsection{Interaction frame without effective fields}\label{sec:effective-frame}
The third approach is similar to the previous one since we include
the rf-field and the chemical shift into an interaction-frame
transformation as a first step, leading to the Hamiltonian in
Eq. (\ref{eq:Hamiltonian-full-frame}).
A repetitive pulse element will introduce a modulation frequency $\omega_\mathrm{m}= 2\pi/\tau_\mathrm{m}$, where $\tau_\mathrm{m}$ is the duration of one basic element. In general,  the direction of a spin before and after a basic element is not identical but rotated, which can be described by a rotation vector $\vec\beta_{\eff}$. The frequency associated by this rotation, is called the effective frequency and defined as $\vec\omega_\eff=\vec\beta_\eff/\tau_\mathrm{m}$. Since the chemical shift has been included in the interaction-frame trajectory the effective frequency is, in general, different for each spin.
In a second step we take the effective field out of the interaction-frame Hamiltonian by applying an interaction-frame transformation with 
\begin{align}
    \UU(t)=\exp{\left(i\, \sum_i \vec\omega^{(i)}_{\eff}\, \vec\II_i\, t\right)}
\end{align}
This transformation introduces the effective fields explicitly in the interaction-frame Hamiltonian, due to the second term of Eq. (\ref{eq:general_transformation}), leading to the interaction-frame Hamiltonian    
\begin{align}
{\HH}(t)
=
\sum_{i,j}
\omega_{ij}(t)
\sum_{\mu,\nu}
A^{(ij)}_{\mu \nu}(t)\, \II_{i\mu} \II_{j\nu} - \sum_i \vec\omega^{(i)}_{\eff}\,\vec\II_i
\end{align}
Notice that the $A^{(ij)}_{\mu\nu}(t)$ are not the same as defined in Eq. (\ref{eq:Hamiltonian-full-frame}). However, the individual interaction-frame trajectories $a_{\mu\nu}(t)$ differ only by a uniform rotation around the effective axis from the trajectories in the complete frame. 
To obtain an effective 
Hamiltonian we transform to the frequency domain and
apply van Vleck perturbation theory to first order resulting in
\begin{align}\label{eq:effective_ham_eff_field}
\bar{\HH}^{(1)}
               & = \sum_{i,j} \sum_n
\omega^{(n)}_{ij} \sum_{\mu,\nu}
\widehat{A}^{(ij)}_{\mu \nu}(n \omega_{r})\, \II_{i\mu} \II_{j\nu}- \sum_i \vec\omega^{(i)}_{\eff}\,\vec\II_i
\end{align}
For simplicity, we can again apply a rotation such that the z axis is aligned with the effective field direction of each
spin as in section \ref{sec:chemical-shift-frame}. This leads to a mixing of the various terms in the effective coupling Hamiltonian, and differing quantization axes for each spin.

%
%
\subsection{Properties of the interaction frames and scaling factors}
In the following, we discuss the three different interaction frames and how to assess the performance of an irradiation
scheme by defining scaling factors for the example of homonuclear dipolar recoupling.
In general, the symmetry of the interaction-frame trajectory, i.e., the
relation between its elements $a_{\mu\nu}$, depends
only on the choice of the basis operators and not on the interaction-frame transformation being used.
A common choice of bases are the Cartesian operators represented by the Hermitian traceless Pauli matrices  $\{\II_x,\II_y,\II_z\}$
where the elements of the frequency-domain interaction-frame trajectory are Hermitian
functions \cite{Chavez:2022} or the basis formed by  $\{\II_+,\II_-,\II_z\}$. 
In the following we use the basis $\{\II_+,\II_-,\II_z\}$, since it
facilitates the understanding of recoupling efficiency in terms of zero-quantum (ZQ) and double-quantum (DQ) transfer more easily, than the Cartesian basis.
In this basis the following relations always hold true
\begin{equation}\label{eq:interaction-frame-symmetry}
\begin{alignedat}{3}
[a^{(ij)}_{++}(t)]^* & =  a^{(ij)}_{--}(t) \quad \quad &&\textrm{or}\quad\quad [\widehat a^{(ij)}_{++}(\Omega)]^*  &&= \widehat  a^{(ij)}_{--}(-\Omega) \\ 
[a^{(ij)}_{+-}(t)]^* & = a^{(ij)}_{-+}(t) \quad \quad &&\textrm{or}\quad\quad  [\widehat a^{(ij)}_{+-}(\Omega)]^* &&= \widehat a^{(ij)}_{-+}(-\Omega) \\
[a^{(ij)}_{z-}(t)]^* & = a^{(ij)}_{z+}(t) \quad \quad &&\textrm{or}\quad\quad [\widehat a^{(ij)}_{z-}(\Omega)]^* &&= \widehat a^{(ij)}_{z+}(-\Omega) \\
[a^{(ij)}_{-z}(t)]^* & = a^{(ij)}_{+z}(t) \quad \quad &&\textrm{or}\quad\quad [\widehat a^{(ij)}_{-z}(\Omega)]^* &&= \widehat a^{(ij)}_{+z}(-\Omega) 
\end{alignedat}
\end{equation}
Notice that complex conjugation flips the sign of the indices as well
as the argument in the case of the frequency-domain interaction-frame
trajectory. Consequently there are at most five independent elements of the interaction-frame trajectory. Here, ``independent'' means that the elements cannot be expressed with each other using complex conjugation and/or inverting the sign of the argument. 
The symmetry of the elements of the interaction-frame trajectory are inherited by the coupling coefficients (see Eq. (\ref{eq:coupling-coefficient})), leading to
\begin{equation}\label{eq:general-symmetry}
\begin{alignedat}{3}
[A^{(ij)}_{++}(t)]^* & =  A^{(ij)}_{--}(t) \quad \quad &&\textrm{or}\quad\quad [\widehat A^{(ij)}_{++}(\Omega)]^*  &&=  \widehat A^{(ij)}_{--}(-\Omega) \\ 
[A^{(ij)}_{+-}(t)]^* & = A^{(ij)}_{-+}(t) \quad \quad &&\textrm{or}\quad\quad  [\widehat A^{(ij)}_{+-}(\Omega)]^* &&= \widehat A^{(ij)}_{-+}(-\Omega) \\
[A^{(ij)}_{z-}(t)]^* & = A^{(ij)}_{z+}(t) \quad \quad &&\textrm{or}\quad\quad [\widehat A^{(ij)}_{z-}(\Omega)]^* &&= \widehat A^{(ij)}_{z+}(-\Omega) \\
[A^{(ij)}_{-z}(t)]^* & = A^{(ij)}_{+z}(t) \quad \quad &&\textrm{or}\quad\quad [\widehat A^{(ij)}_{-z}(\Omega)]^* &&= \widehat A^{(ij)}_{+z}(-\Omega) 
\end{alignedat}
\end{equation}
Therefore there are at most 5 independent coupling coefficients.
However, if the two spins have the same interaction-frame trajectory 
as in the interaction frame generated by the radio-frequency Hamiltonian (see section \ref{sec:chemical-shift-frame}), we have in addition 
\begin{align}
    A_{\mu\nu}(t) =  A_{\nu\mu}(t) \quad \quad \textrm{or}\quad\quad  \widehat A_{\mu\nu}(\Omega) =  \widehat A_{\nu\mu}(\Omega)
\end{align}
which reduces the number of independent coefficients to four. 

%
%
\subsubsection{Properties of the interaction frame without chemical shift}
\label{sec:csinteractionframe}
As mentioned before, this frame is particularly useful to understand the role of the
chemical-shift offset, since it is not contained in the frame. The additional chemical-shift term in the Hamiltonian is solely responsible for any chemical-shift-dependent
effect since the modulated coupling term will be the same for all spin pairs. In general, there are three components per spin, which are
proportional to $\omega_i \widehat{a}_{z+}(0), \omega_i
\widehat{a}_{z-}(0)$ and $\omega_i \widehat{a}_{zz}(0)$. The direction defined by this vector is the same for every spin, whereas the length of the vector depends on the spin.  However, in many pulse schemes with high enough symmetry with respect to the z direction $\widehat{a}_{zx}(0) \approx 0 \approx \widehat{a}_{zy}(0)$ and, therefore, we can assume that the chemical-shift term
has only a $\II_{z}$ component. In the following, we focus on
this case, but the general case when $\widehat{a}_{zx}(0) \not \approx 0 \not \approx \widehat{a}_{zy}(0)$ can be found in the SI.
One possibility to quantify the effect of the chemical shift on the resonant coupling is via an
additional interaction-frame transformation with the
scaled chemical-shift term, leading to
\begin{align}\label{eq:first-order-CS}
  \bar{\HH}^{(1)}(t)
  = \sum_{i,j} \sum_{\mu,\nu}\sum_{n} \omega^{(n)}_{ij}
  & \,\widehat{A}_{\mu\nu}(n \omega_{\rr})\,
    \II_{i\mu} \II_{j\nu}\,
    \exp \left(i\,\frac{\widehat{a}_{zz}(0)\, \Theta^{(ij)}_{\mu\nu}  t}{2}\right)
\end{align}
with
\begin{align}
 \mathbf{\Theta}^{(ij)}= \left(
  \begin{matrix}
     \omega_i+\omega_j       & \omega_i-\omega_j        & \omega_i        \\
    -\omega_i+\omega_i       & -\omega_i-\omega_j       & -\omega_i       \\
    \omega_j                       & -\omega_j                      & 0
  \end{matrix}
  \right)
\end{align}
In Eq. (\ref{eq:first-order-CS}) we see, that the truncation effect is caused
by an averaging effect due to the phase factor $\exp(i\,\widehat a_{zz}(0)\Theta^{(ij)}_{\mu\nu}t)$.
To see this effect more clearly, we
average the Hamiltonian over the duration of the irradiation, leading to
\begin{align}
 \frac{1}{T}\int_{-T/2}^{T/2}\bar{\HH}^{(1)}(t) dt
  = \sum_{\mu,\nu}\sum_{n}\sum_{i,j} \omega^{(n)}_{ij}
                                         & \,\widehat A_{\mu\nu}(n\omega_\rr)\,\II_{i\mu} \II_{j\nu}\, \textrm{sinc}\left({\frac{\widehat a_{zz}(0)\, \Theta^{(ij)}_{\mu\nu}  T}{2}}\right)
\end{align}
where $\textrm{sinc}(x)=\sin(x)/x$.
Notice that for $\widehat{a}_{zz}(0)=0$ no truncation of the resonant dipolar-coupling term occurs, neither in the ZQ nor in the DQ term. Therefore, every broadband recoupling pulse scheme has to generate $\widehat{a}_{zz}(0)\approx0$.  
If $\widehat{a}_{zz}(0) \neq 0$, the DQ and the ZQ terms will be truncated whenever $\omega_i\pm\omega_j = 0$ is not fulfilled, respectively. The strength of the truncation solely depends on the size of $|\text{sinc} (\widehat a_{zz}(0) \Theta_{\mu\nu}^{(ij)} T)|$; the bigger this term, the stronger the truncation of the corresponding ZQ or DQ term in the Hamiltonian. Notice, that the duration of the irradiation also contributes to the strength of the truncation.
Therefore a sensible definition of the scaling factor, which takes the truncation due to the chemical shift into account, is given by
\begin{align}\label{eq:scaling-factor-cs}
 C_{\mu\nu}^{(ij)} = \,Q_{\mu\nu}^{(ij)} \sum^{2}_{n=-2}  \left|\widehat{A}_{\mu\nu}(n\omega_\rr)\,\right|^2
\end{align}
where we introduced a truncation factor as 
\begin{align}\label{eq:truncation-factor-cs}
Q_{\mu\nu}^{(ij)}=\left|\textrm{sinc}\left(\frac{\widehat{a}_{zz}(0)\, \Theta^{(ij)}_{\mu\nu}  T}{2}\right)\right|^2 
\end{align}
\begin{figure}[h]
\includegraphics[width=0.9\textwidth]{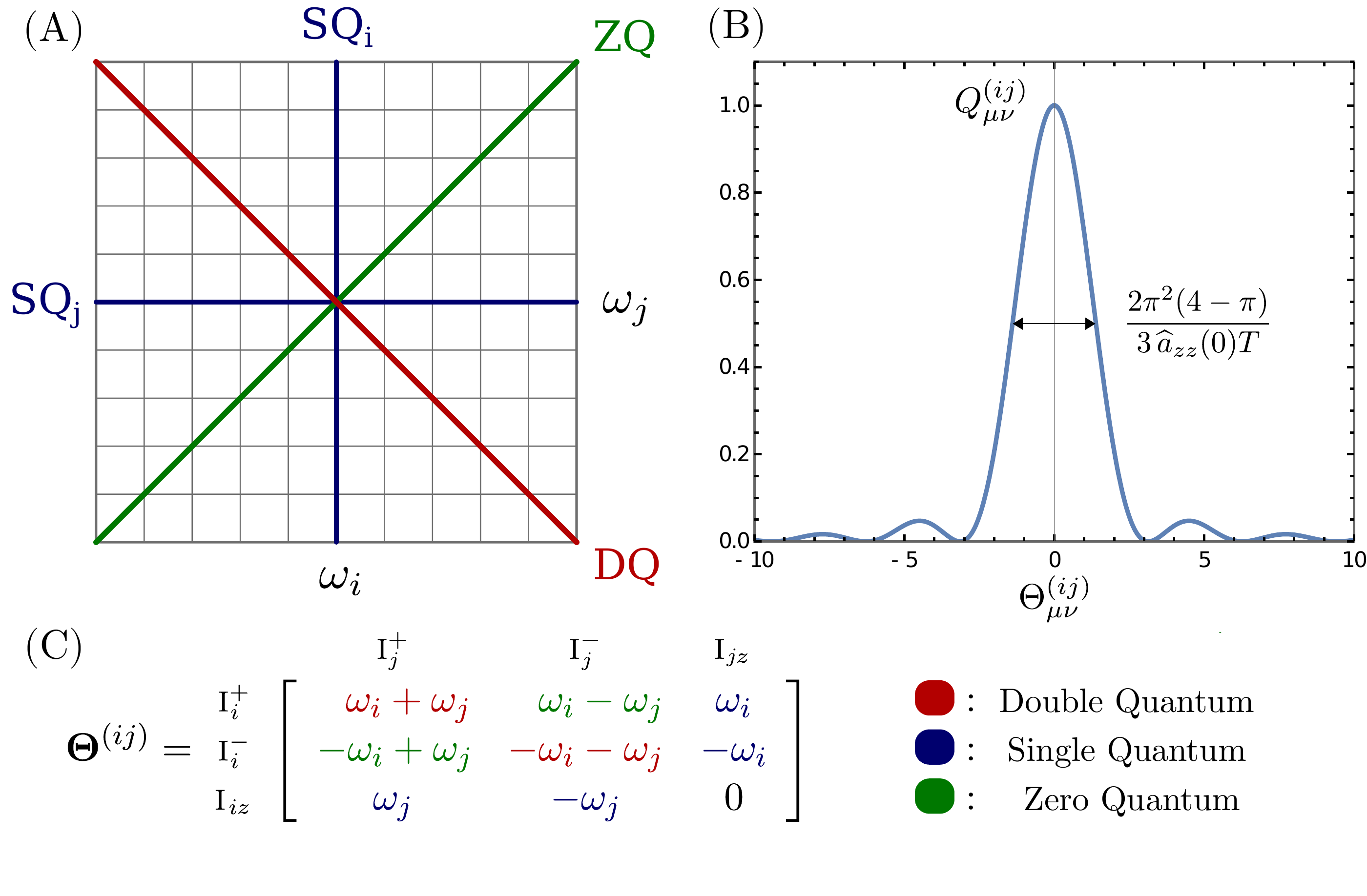}
\caption{(A) Illustration of the conditions $Q^{(ij)}_{\mu\nu} = 1$, hence $\Theta^{(ij)}_{\mu\nu}=0$, which are not truncated by the chemical shifts (blue $z\pm$ (SQ), black $\pm z$ (SQ), red $\pm\pm$ (DQ), green $\pm\mp$ (ZQ)). (B) plot of $Q^{(ij)}_{\mu\nu}$ as a function of $\Theta^{(ij)}_{\mu\nu}$ for $\widehat{a}_{zz}(0)T=2$. The FWHM of $Q^{(ij)}_{\mu\nu}$ is indicated in the plot and given by $\frac{2\pi^2(4-\pi)}{3\,\widehat{a}_{zz}(0)T}$. Therefore, the larger $\widehat{a}_{zz}(0)T$ the narrower the potential recoupling bandwidth around $\Theta^{(ij)}_{\mu\nu}=0$.  The elements of the matrix $\Theta^{(ij)}_{\mu\nu}$ are given in (C).}
\label{fig:Q}
\end{figure}
The matrix $\mathbf{C}^{(ij)}$ has the structure as shown in Eq. (\ref{C-matrix}), consisting of at most 5 different elements. The function
$Q_{\mu\nu}^{(ij)}$ determines the suppression of the recoupling due to the isotropic chemical shifts.
This is illustrated in Figure \ref{fig:Q} (A) which depicts the conditions $\Theta^{(ij)}_{\mu\nu}=0$ which leads to  $Q_{\mu\nu}^{(ij)}=1$ for different combinations of $\mu\nu$ along the colored lines (blue $z\pm$ (SQ), black $\pm z$ (SQ), red $\pm\pm$ (DQ), green $\pm\mp$ (ZQ)) and corresponds to the chemical-shift combinations without truncation.
For example, if $\left|\Theta^{(ij)}_{++}\right|=0$ then $Q_{++}^{(ij)}=1$ independent of  $\widehat{a}_{zz}(0)$, hence the DQ recoupling given by $\left|\widehat A_{++}(n\omega_\rr)\,\right|^2$ will not be truncated on the red line where the two chemical shifts have the same magnitude but opposite sign.
The full width at half maximum (FWHM) and roots of $Q_{\mu\nu}^{(ij)}$ are given by 
\begin{align}
    \Delta\omega \approx \frac{2\pi^2(4-\pi)}{3\,\widehat{a}_{zz}(0\,) T} \label{eq:FWHM},
\hspace{0.5cm}
    \omega_0 = \frac{2\pi\, n}{\widehat{a}_{zz}(0)\,T} \hspace{0.5cm} n\in\mathbb{Z} 
\end{align}
respectively, where the FWHM was approximated with the sine approximation formula by Bhaskara I.
Notice that the FWHM of $Q_{\mu\nu}^{(ij)}$, and, therefore, the range of chemical-shift combinations which are not strongly truncated, is inversely proportional to $\widehat{a}_{zz}(0)T$ (see Fig. \ref{fig:Q} (B)).
\begin{figure}
\centering
\includegraphics[width=0.5\textwidth]{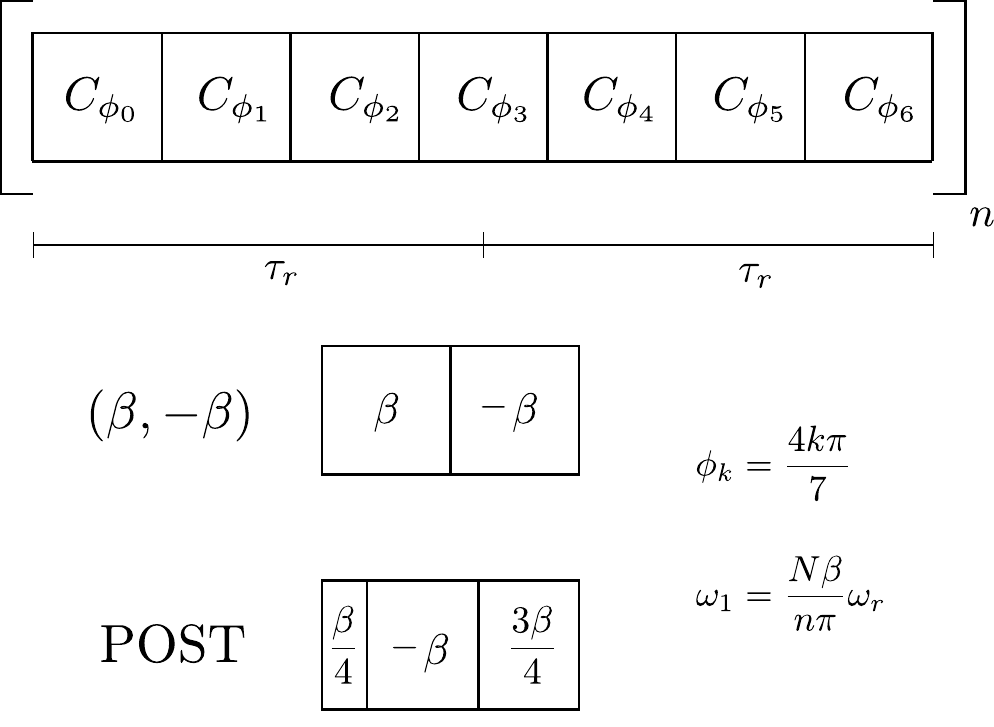}
\caption{Modification of the $C7_2^{1}$ pulse scheme for selective recoupling. As the basic element we can either use $(\beta,-\beta)_{\phi}$ or a post element $(\beta/4,-\beta,3\beta/4)_\phi$. The minus sign means that the rotation is in the opposite direction, i. e. a phase shift of $\pi \, \text{rad}$. Notice that the rf-field amplitude depends on the flip angle $\beta$, but the duration of the scheme stays the same.}
\label{fig:sequences}
\end{figure}
%
%
\subsubsection{Properties of the complete interaction frame}
\label{sec:interpretation-complete}
In the complete interaction frame every spin with a different chemical shift has its
individual frame and, therefore, the aggregated interaction frame
transformation has five independent coefficients.
$\widehat A^{(ij)}_{\mu\nu}(\Omega)$ encodes all the information of the
irradiation and of the chemical shift. In addition, it directly scales the
two-spin operators and, therefore, the effective dipolar coupling under the pulse sequence. As a result, we can directly characterize the effect of an irradiation scheme for a specific spin pair and get insight into the prevailing transfer mechanism (DQ, ZQ). One possibility to define such a scaling factor from  $\widehat A^{(ij)}_{\mu\nu}(\Omega)$ is 
\begin{align}\label{eq:scaling-factor-complete}
C_{\mu \nu}^{(ij)} = \sum_{n=-2}^{2} |\widehat A^{(ij)}_{\mu \nu}(n \omega_\textrm{r})|^2
\end{align}
This quantity can be interpreted as the magnitude of the resonance conditions given by the interplay between the spin modulation, i.e. the irradiation and the chemical shift, and the spatial  modulation, i.e., the magic-angle spinning. Since $C_{\mu \nu}^{(ij)}$ is defined with an absolute square and a sum over a symmetric interval around 0, it has at most five different elements:
\begin{align}
  \label{C-matrix}
\mathbf{C}^{(ij)}
=
\begin{pmatrix}
C_\mathrm{DQ}                                   & C_\mathrm{ZQ}                               & C_\mathrm{SQ_1}              \\
C_\mathrm{ZQ}                                   & C_\mathrm{DQ}                               & C_\mathrm{SQ_1}              \\
C_\mathrm{SQ_2}                                 & C_\mathrm{SQ_2}                             & C_\mathrm{zz}
\end{pmatrix}
\end{align}
The second major benefit of this frame, besides simplicity, is that it
leads to the maximal reduction of the energy splittings in the Hamiltonian and, therefore, the
fastest convergence of the perturbation series, compared to other
frames.  On the other hand, this frame does not offer any analytical understanding, neither of the
role of the chemical shifts, nor of the effective field, because they
are both contained in the modulations of the coupling term.

The features of this interaction frame makes it
particular suitable for any application, where we want a simple and computationally efficient way to
characterize the performance of a pulse scheme, including its
chemical-shift offset dependence. For example, it is the most promising
frame for the definition of the cost function for an optimization procedure.
%
%
\subsubsection{Properties of the interaction frame without effective fields}
\label{sec:interpretation-eff}
The effective field frame is also an individual frame for each spin in the same way as the complete
interaction frame (see section \ref{sec:interpretation-complete}). In contrast to the chemical-shift interaction frame, the additional term, i.e., the effective field
term, is time independent in the interaction frame and can be taken out. If we again consider the case where
$\omega_i^{(\eff)}$ is close to the z axis, we can use the same procedure as in section \ref{sec:csinteractionframe}. The general treatment can be found in the SI. We perform an additional interaction-frame transformation with the effective-field Hamiltonian on the first-order effective Hamiltonian and obtain a phase modulation of the coupling elements where the
phase matrix is
\begin{align}\label{eq:effective-field-theta}
\mathbf{\Theta}^{(ij)}= \left(
  \begin{matrix}
     \omega^{(\eff)}_i+\omega^{(\eff)}_j & \omega^{(\eff)}_i-\omega^{(\eff)}_j  & \omega^{(\eff)}_i     \\
    -\omega^{(\eff)}_i+\omega^{(\eff)}_i & -\omega^{(\eff)}_i-\omega^{(\eff)}_j & -\omega^{(\eff)}_i    \\
    \omega^{(\eff)}_j                    & -\omega^{(\eff)}_j                   & 0
  \end{matrix}
\right)
\end{align}
This leads again to a scaling factor due to the truncation of the coupling terms by the effective field which is given by
\begin{align}\label{eq:scaling-factor-eff}
 C_{\mu\nu}^{(ij)} = \,Q_{\mu\nu}^{(ij)} \sum^{2}_{n=-2}  \left|A^{(ij)}_{\mu\nu}(n\omega_\rr)\,\right|^2
\end{align}
where we introduced the truncation factor as 
\begin{align}\label{eq:truncation-factor-eff}
Q_{\mu\nu}^{(ij)}=\left|\textrm{sinc}\left(\frac{\Theta^{(ij)}_{\mu\nu}  T}{2}\right)\right|^2 
\end{align}
The more general case where the two effective fields are not along the z axis and have possibly different directions in space, requires first a coordinate transformation such that the effective fields are along the z axis. This will lead to a mixing of the various DQ, ZQ, and SQ terms in the effective Hamiltonian of Eq. (\ref{eq:effective_ham_eff_field}). We will not discuss this case in more detail here, but it can be found in the SI.
%
%
\section{Application to symmetry-based C sequences}
\begin{figure}
\includegraphics[width=\textwidth]{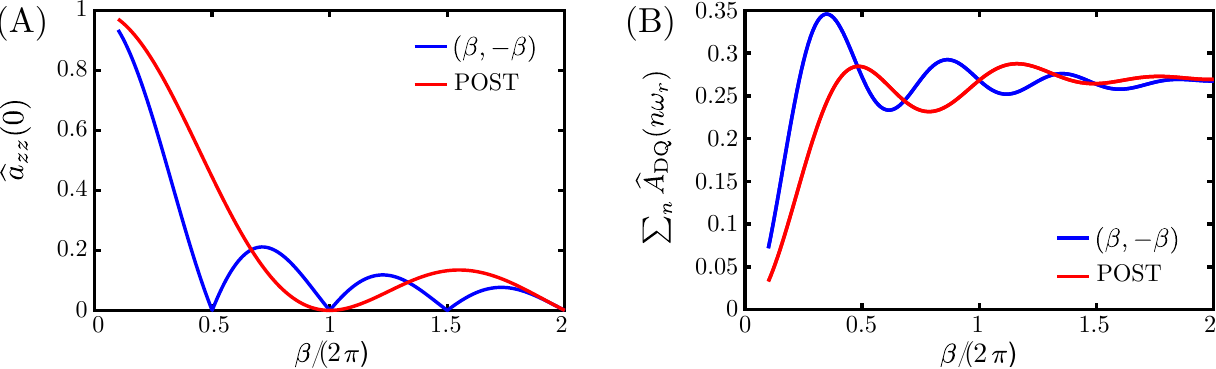}
\caption{Plot of $\widehat{a}_{zz}(0)$ and $\sum_n\widehat A_{\text{DQ}}(n\omega_\rr)$ for the pulse scheme of Fig. \ref{fig:sequences} as function of the flip angle $\beta$. (A) The larger $\widehat{a}_{zz}(0)$ the stronger the truncation due to the chemical shift, according to Eq. (\ref{eq:truncation-factor-cs}). For $\beta \approx \pi$ the polarization transfer is only truncated weakly in the case of the $(\beta,-\beta)_{\phi}$ element, however truncated strongly when using a POST element. As a result, the $(\beta,-\beta)_{\phi}$ element allows for a significant reduction of the required rf-field amplitude compared to the POST element.
(B) Both basic elements lead to a similar $\sum_n\widehat A_{\text{DQ}}(n\omega_\rr)$ and therefore to a similar maximal transfer efficiency, which is not changed significantly in the relevant range for $\beta$.}
\label{fig:coefficients}
\end{figure}

\begin{figure}
  \centering
  \includegraphics[width=1\textwidth]{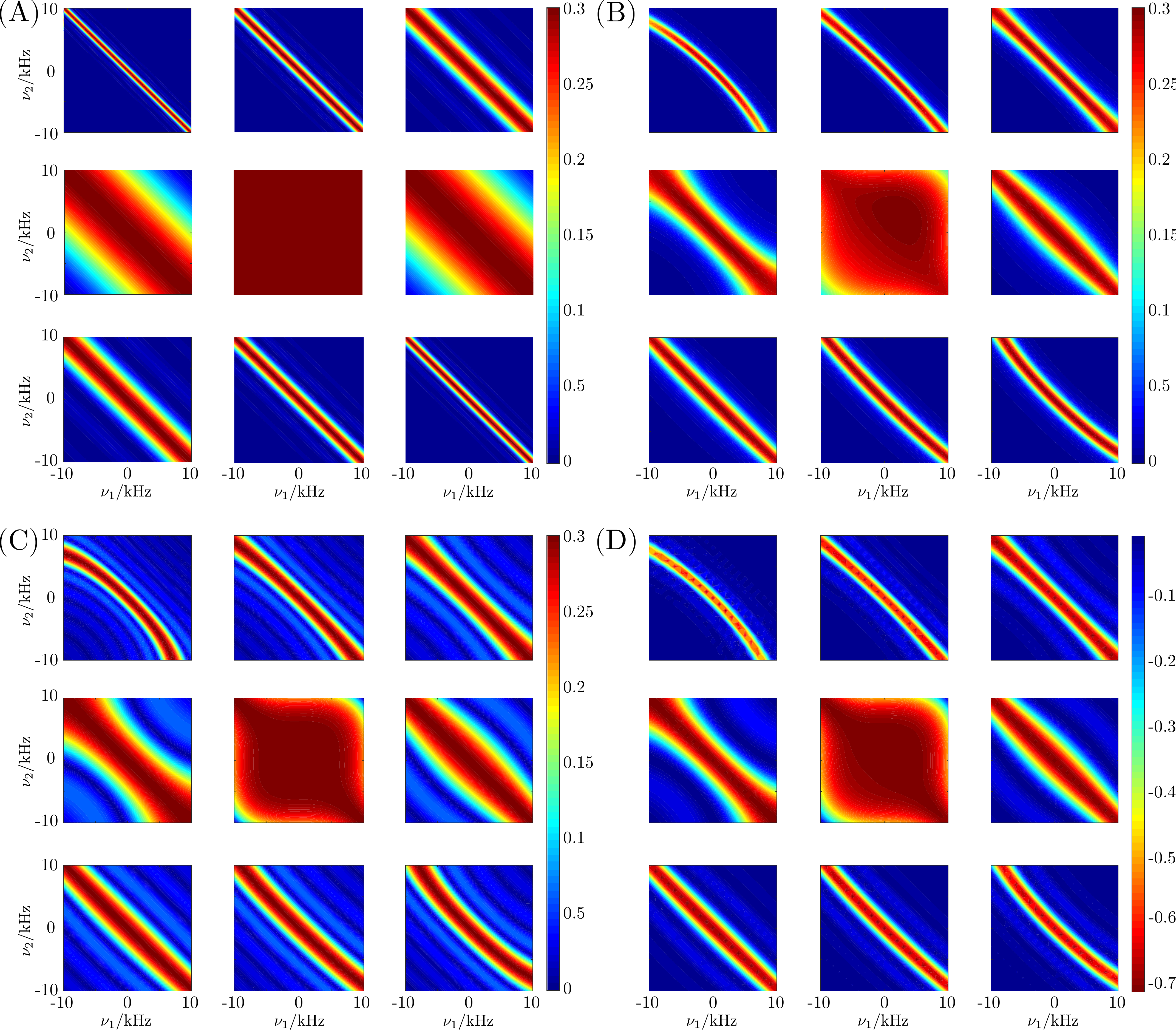}
  \caption{Scaling factor $C^{(ij)}_{\text{DQ}}$ in the interaction frames: (A) without chemical shift  (Eq. (\ref{eq:scaling-factor-cs})), (B) complete (Eq. (\ref{eq:scaling-factor-complete})), (C) without effective fields (Eq. (\ref{eq:scaling-factor-eff})).
Subplot (D) shows the transfer efficiency  obtained by numerical time-slicing simulations.
The quantities are displayed as a function of the chemical shifts of a homonuclear two-spin system for the pulse scheme shown in Fig. (\ref{fig:sequences}) with the $(\beta,-\beta)_{\phi}$ basic element.
In each grouping $\beta$ ranges from $280^{\circ}$ to $440^{\circ}$ in steps of $20^{\circ}$.}
\label{fig:comparison}
\end{figure}

Symmetry based $\mathrm{C}N_n^\nu$ sequences \cite{Levitt:2007tg} can be used as mixing sequences for broadband double-quantum recoupling experiments. The chemical-shift compensation, i.e., the scaling  $\widehat{a}_{zz}(0) \approx 0$, is achieved by the basic building block of the sequence and not the phase rotation of the sequence. The basic building block can be any element which generates an identity propagator and is typically either a $(2\pi,-2\pi)$ element \cite{Nielsen:1994tz} or a so-called POST $(\pi/2,-2\pi,3\pi/2)_\phi$ element \cite{Hohwy:1998vv}.
The POST element ensures a more robust chemical-shift compensation over a larger range of chemical shifts, which corresponds to $\widehat{a}_{zz}(0) \approx 0$ for larger chemical-shift values.
\\
The scaling factor of the dipolar recoupling characterized by $\left|A^{(ij)}_{\mu\nu}(n\omega_\rr)\,\right|^2$ is not strongly influenced by the choice of the basic building block as long as the basic-building block generates an identity operation, i.e., the direction of the interaction frame is the same before and after the basic element. As a consequence we can modify the basic building block of $\mathrm{C}N_n^\nu$ sequences to make $\widehat{a}_{zz}(0)$ nonzero without compromising the recoupling efficiency significantly. Such a modified C sequence is depicted in Figure \ref{fig:sequences}, where the basic building block of the $\mathrm{C}7^1_2$ pulse sequence is modified. As shown in the Figure \ref{fig:coefficients}, $\widehat{a}_{zz}(0)$ can be tuned by changing the flip angle $\beta$ without influencing  $\left|A^{(ij)}_{\mu\nu}(n\omega_\rr)\,\right|^2$ significantly.
\\
This principle has been used in a series of selective dipolar recoupling sequences  \cite{Agarwal:2018kc, Potnuru:2020ez, Duong:2020gc, Zhang:2020fq, Potnuru:2021cn, Xiao:2021ih} where either a numerical optimization or a modification of existing recoupling sequences was used to generate sequences that have good recoupling efficiency and poor chemical-shift compensation. However, the theoretical basis of these sequences was not always clearly stated in the publications. One of the proposed selective recoupling sequences uses explicitly the modification of the $\mathrm{C}N_n^\nu$  sequences as discussed above \cite{Potnuru:2021cn}. Here, we use the analysis and design of these type of pulse schemes as a case study to illustrate the presented techniques and interaction frames.
\\
Since our interest lies mainly in the chemical-shift compensation of the $\mathrm{C}7^1_2$ pulse scheme, we begin our analysis in the interaction frame presented in section (\ref{sec:chemical-shift-frame}) and (\ref{sec:csinteractionframe}), where the influence of the chemical shift  is isolated by keeping it in the Hamiltonian. The recoupling strength in this interaction frame is described by Eq. (\ref{eq:scaling-factor-cs}) that contains two quantities: (i) the dipolar scaling factor $\widehat A_{\mu\nu}(n\omega_\rr)$ which determines the maximal transfer efficiency and (ii) the scaling factor of the chemical-shift term $\widehat a_{zz}(0)$ that modulates the transfer efficiency through the $Q_{\mu\nu}^{(ij)}$ parameter (Eq. (\ref{eq:truncation-factor-cs})), which describes the truncation of the effective dipolar coupling by the effective chemical shift. For the physical intuitive understanding of the working of such sequences and simple discussions, the interaction frame generated by the radio-frequency irradiation alone provides the best framework.
\\
Figure \ref{fig:sequences} (A) shows a simple modification of the $\mathrm{C}7^1_2$ double-quantum recoupling sequence \cite{Potnuru:2021cn} where the basic building block was modified from $(2\pi,-2\pi)_\phi$  to $(\beta, -\beta)_\phi$. The same principle can also be implemented for the POST basic element by going from $(\pi/2,-2\pi,3\pi/2)_\phi$ to $(\beta/4,-\beta, 3\beta/4)_\phi$. Calculating the two scaling factors on the resonance condition as a function of the rf-field amplitude will allow us to judge the performance of the $\mathrm{C}7^1_2$ sequence.
\\
Figure  \ref{fig:coefficients} (A) shows the $\widehat A_{\text{DQ}}(n\omega_\rr)$ scaling factor as a function of the effective flip angle $\beta$, where $\beta=2\pi$ would correspond to the ideal $\mathrm{C}7^1_2$ sequence. One can clearly see that over a large range of $\beta$ values, the dipolar scaling factor does not change significantly. On the other hand Figure \ref{fig:coefficients} (B) shows the chemical-shift scaling factor $\widehat a_{zz}(0)$ which changes significantly with the value of $\beta$. For values of $\beta=2\pi$ and $\beta=\pi$ and the $(\beta,-\beta)_{\phi}$ basic element, the chemical-shift scaling factor is zero, leading to a broadband dipolar recoupling sequence in first order. Using the POST element, only $\beta=2\pi$ achieves broadband recoupling.
\\
We can now use the scaling factor of Eq. (\ref{eq:truncation-factor-cs}) to tune the bandwidth of the dipolar recoupling sequence by changing the value of $\beta$, which is exactly what has been done in \cite{Potnuru:2021cn}. For a DQ recoupling sequence, we can use Eq. (\ref{eq:scaling-factor-cs}) as a measure for the recoupling efficiency as a function of the two chemical-shift offsets (Figure \ref{fig:comparison} (A)) for different values of the flip angle $\beta$. For $\beta=2\pi$ (center plot), broadband recoupling is achieved because the effective fields are small over the full range of chemical shifts while for other values of $\beta$ only a band-selective recoupling is achieved where the width depends on the flip angle $\beta$.
As a simpler but equivalent measure, one could also use directly the sum of the two scaled chemical shifts ($a_{zz}(0)(\omega_i+\omega_j)$) that truncate the effective dipolar coupling term (Eq. \ref{eq:truncation-factor-cs}), as a measure for the recoupling efficiency. The corresponding plots can be found in the SI.
\\
Figure \ref{fig:comparison} (B) shows the DQ scaling factor in the complete interaction frame (Eq. (\ref{eq:scaling-factor-complete})) which gives little intuitive insight into the properties of the sequence. On the other hand, there is only a single parameter to evaluate which makes this interaction frame interesting and most suitable for numerical optimization of sequences since the convergence of the effective Hamiltonian series, especially for larger chemical-shift values, is faster than in the previous case.
\\
On the other hand, the interaction frame without effective field is characterized by Eq. (\ref{eq:scaling-factor-eff}) and is shown in Fig. \ref{fig:comparison} (C). In this case, we can also use the magnitude of the effective fields as a simplified measure of how strong the effective dipolar coupling is truncated in a way similar to the first case discussed, as shown in the SI. These effective fields have been used as a measure for solution-state TOCSY efficiency \cite{Nielsen:2019kf} as a function of the chemical-shift offset.
\\
Figure \ref{fig:comparison} (D) shows the efficiency of the polarization transfer using exact numerical simulations in a two-spin system as a function of the chemical-shift offset for different flip angles $\beta$. The simulations show the maximum of the polarization transfer for a fixed mixing time of $2\,$ms and a dipolar coupling of $\delta/(2\pi) = -2425\,$Hz. The overall shape of the numerical simulation is in good agreement with the predictions based on simple interaction-frame pictures.
The most striking difference is the deviation of the optimum transfer from the anti diagonal in the numerical simulations from the scaling factor in the interaction frame without chemical shift. 
However the complete interaction frame and the interaction frame without effective field hardly deviate from the numerical simulations.

\begin{figure}[h]
  \includegraphics[width=0.7\textwidth]{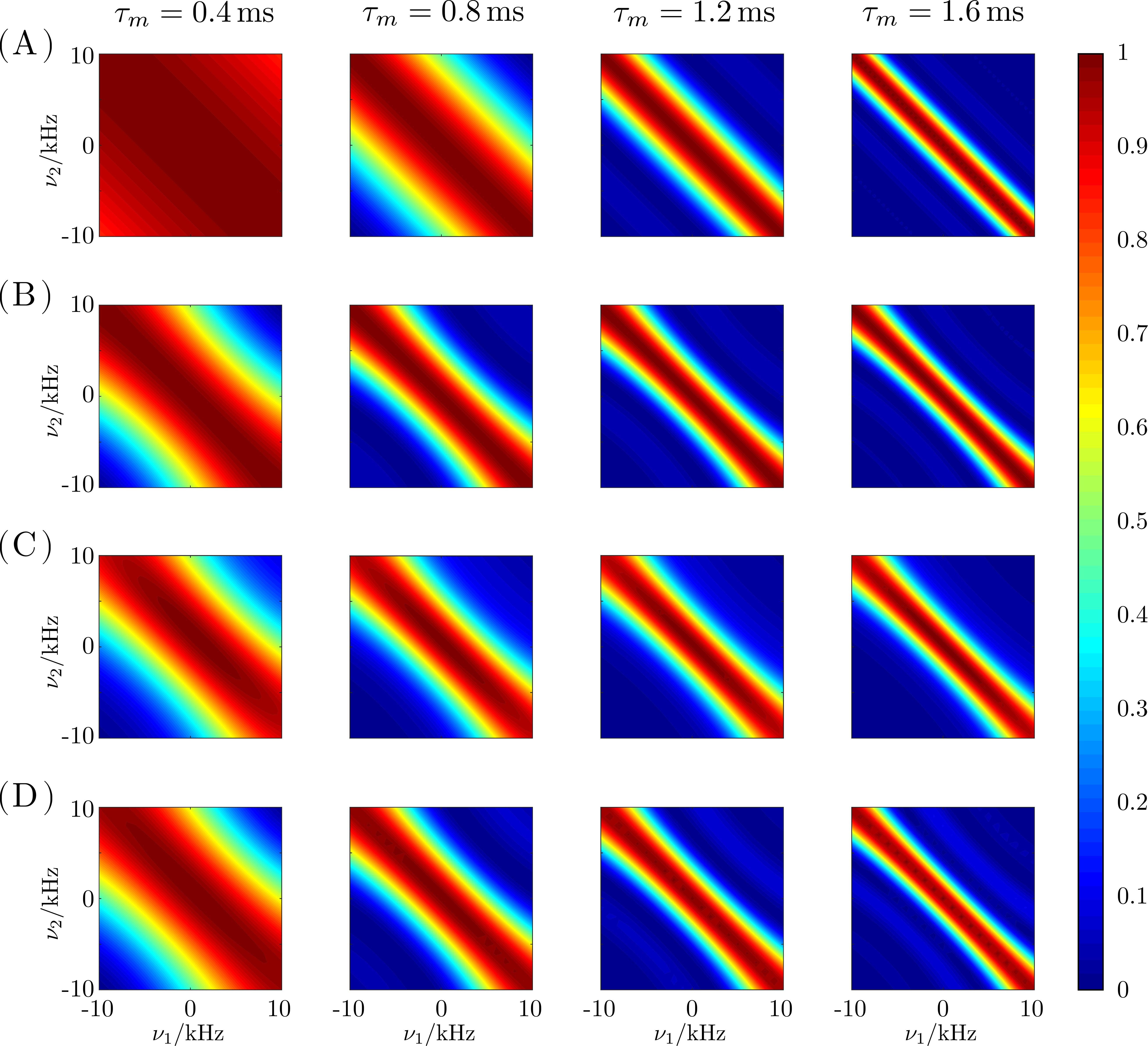}
  \caption{Scaling factor $C^{(ij)}_{\text{DQ}}$ in the interaction frames: (A) without chemical shift  (Eq. (\ref{eq:scaling-factor-cs})), (B) complete (Eq. (\ref{eq:scaling-factor-complete})), (C) without effective fields (Eq. (\ref{eq:scaling-factor-eff})) for the same $C7^1_2$ modification with $\beta=320^\circ$ but different mixing times. Subplot (D) shows the transfer efficiency obtained by numerical simulations. The plots in a row are normalized to 1 among themselves. The mixing time $\tau_\mathrm{m}$ ranges form $0.4\,$ms to $1.6\,$ms in steps of $0.4\,$ms. As expected we see a linear decrease of the FWHM of the recoupling bandwidth with increasing mixing times.}
  \label{fig:mixing_time}
\end{figure}

Figure \ref{fig:mixing_time} illustrates the dependence of the recoupling bandwidth for the implementation with $\beta = 320^\circ$ on the mixing time $\tau_\mathrm{m}$. Figure \ref{fig:mixing_time} (A) and (B) show the scaling factor of the interaction frame without chemical shift and without effective field given by Eq. (\ref{eq:scaling-factor-cs}) and Eq. (\ref{eq:scaling-factor-eff}), respectively. 
The expected linear dependence between the recoupling bandwidth (FWHM) and mixing time $\tau_\mathrm{m}$ are in good agreement with the  numerical simulations shown in Fig. \ref{fig:mixing_time} (D).
The narrowing of the recoupling bandwidth for longer mixing times has been observed between SPR-$5_6$ and SPR-$5_4$ and reported in \cite{Zhang:2020fq}. 
The observed decrease from $\sim0.3\,$kHz to $\sim0.5\,$kHz in the recoupling bandwidth between SPR-$5_6$ and SPR-$5_4$ can be accurately explained by the different mixing times, $\tau_\mathrm{m} \approx 3.2\,$ms and $\tau_\mathrm{m}\approx 1.8\,$ms, used.

Notice that the dependence of the mixing time is explicitly introduced by the additional interaction-frame transformation in the case of the interaction frames without chemical shift and without effective field. However in the complete interaction frame the dependence on the mixing time is implicitly encoded in the coupling coefficients $A^{(ij)}_{\mu\nu}(n\omega_\rr)$. 
%
%
\section{Conclusion}
In this paper we discuss the use of interaction-frame transformations in the context of continuous Floquet theory for solid-state NMR.
Using the example of chemical-shift-selective recoupling with symmetry-based sequences, we illustrate how different interaction-frame transformations can be used for different situations and to answer different questions.
By excluding a specific process or interaction from an interaction-frame transformation, its effect can be isolated and, therefore, understood in a simpler analytical way.
As presented in section \ref{sec:chemical-shift-frame} and \ref{sec:effective-frame}, the terms of interest can be excluded from the interaction-frame transformation in the beginning, or excluded with a subsequent second transformation, respectively. However the first approach is generally only possible, if the isolated interactions are explicitly given in the Hamiltonian, otherwise the second technique has to be used.  
In contrast to the interaction frames that isolate a specific term, all the single-spin terms can be included in the interaction frame. The Hamiltonian in this frame will always only contain two-spin terms, reducing the complexity significantly.
Such an approach does not offer analytical insights of any particular effect, but leads directly to a scaling coefficient for the coupling strength which includes all the contributions encoded in the Hamiltonian. In the full interaction frame, the perturbation series convergence faster than in interaction frames, that do not include the complete single-spin dynamics. In addition, calculating higher-order terms in this interaction frame is more convenient, because of the simplicity of the Hamiltonian. 
\section*{Declaration of competing interest}
The authors have no conflicts of interest to disclose.

\section*{Data availability}
The simulated data will be uploaded to a public repository after acceptance of the paper.

\section*{Acknowledgements}
This research has been supported by the ETH Zürich and 
the Schweizerischer Nationalfonds zur Förderung der Wissenschaftlichen
Forschung (grant no. 200020\_188988). 

\bibliography{library.bib}
\end{document}


\bibliographystyle{unsrt}
\title{Supplementary Information: Interaction frames in solid-state NMR: A case study for chemical shift selective irradiation schemes}

\author[1]{Matías Chávez}
\affil{\small Physical Chemistry, ETH Zürich, Vladimir-Prelog-Weg 2, 8093 Zürich, Switzerland}

\author[1]{Matthias Ernst}%
\maketitle

\section{Chemical-shift and effective-field frequencies}
\begin{figure}[h]
  \includegraphics[width=\textwidth]{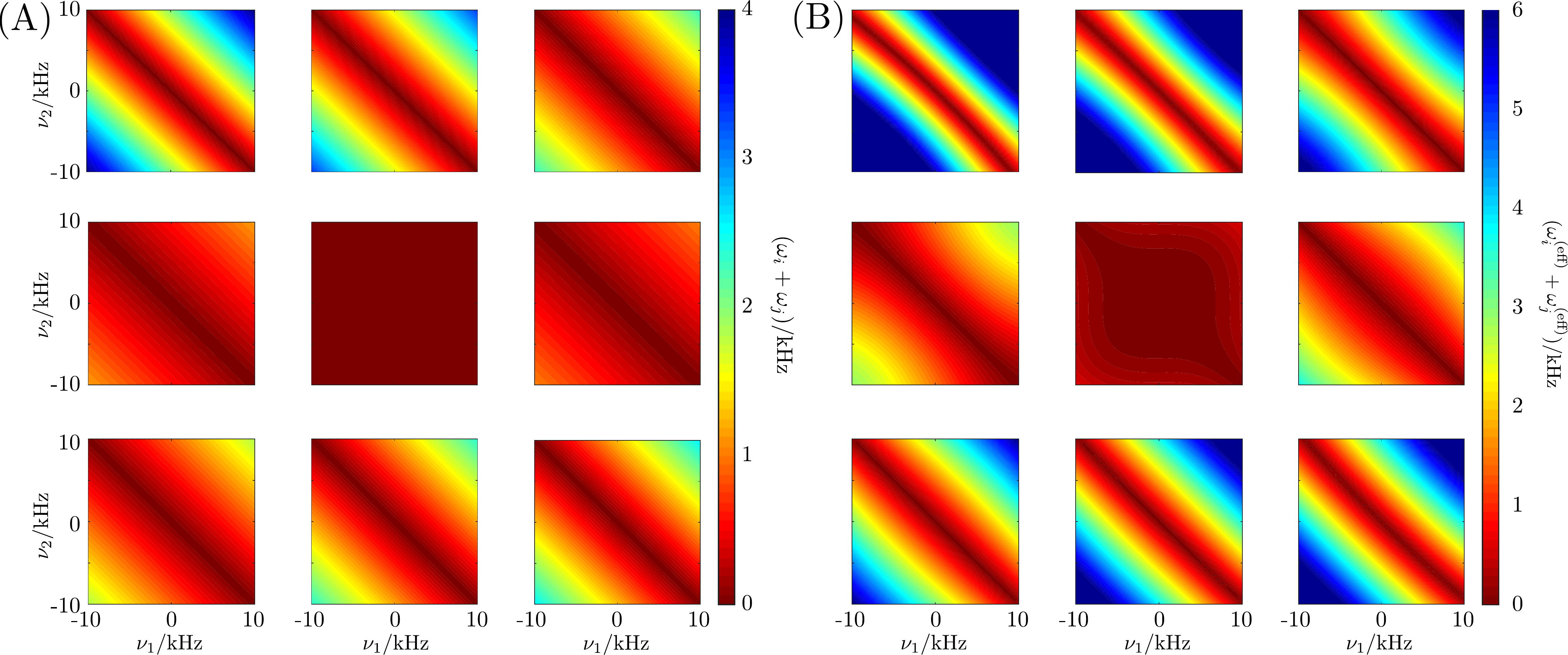}
  \caption{(A) Sum of the chemical shift frequency as function of the chemical shift offset. (B) Sum of the effective-field frequencies as function of the chemical shift offset. Both quantities give an qualitative estimate of the truncation effect.}
  \label{fig:fields}
\end{figure}

Figure \ref{fig:fields} shows the sum of chemical shift frequencies and the sum of the effective-field frequencies as function of the chemical shift offset. As mentioned in the main text they can be used to give an estimate of the truncation effect. However they are in general less precise then the truncation factor $Q_{\mu\nu}$ and do not describe the effect of the mixing time on the truncation effect.

\section{Calculation of the scaling factors in the general case}

\subsection{Interaction frame without chemical shift}

In this section we consider the general case, where  $\widehat{a}_{zx}(0) \not \approx 0 \not \approx \widehat{a}_{zy}(0)$, for the Hamiltonian of section 2.2.1 in the main text
\begin{align}\label{eq:HamiltonianCS}
\bar{\HH}^{(1)} = \sum_n {\HH}^{(n)}(n\omega_\textrm{r})
=
\sum_n
\sum_{i,j}
\omega^{(n)}_{ij}\sum_{\mu,\nu}\widehat{A}_{\mu \nu}(n\omega_\textrm{r})\, \II_{i\mu} \II_{j\nu}
+ \sum_{i=1}^N\sum_{\chi}\omega_{i} \widehat{a}_{z\chi}(0) \, \II_{i\chi}
\end{align}
In this case we can align the effective chemical-shift vector $\vec a_{z}(0) = (\widehat{a}_{z,x}(0), \widehat{a}_{z,y}(0), \widehat{a}_{z,z}(0))^T$ and the z axis by a (immediate) rotation $R_{\mu\nu}$. This leads to the rotated vector $\vec {\tilde{a}}_{z}(0) = (0,0, \widehat{\tilde{a}}_{z,z}(0))^T$, where we use the tilde to differentiate the rotated and the non-rotated vector.
Now we can apply the same formalism as described in section 2.2.1 and obtain the scaling factor $\tilde{C}_{\mu\nu}$.
However, the rotation  $R_{\mu\nu}$ lead to an mixing of the components of $\widehat{A}_{\mu \nu}(\Omega)$. To account for this mixing we have to apply the inverse rotation on the scaling factor as
\begin{align}
  \label{eq:inverse_rotation}
  C_{\mu\nu}=\sum_{\chi\chi'} R^{(-1)}_{\chi\mu} R^{(-1)}_{\chi'\nu} \tilde{C}_{\chi\chi'}
\end{align}
As a result we obtain the scaling factor $C_{\mu\nu}$ in the usual spin basis, which can be interpreted as described in the main text.

\subsection{Interaction frame without effective field}
A similar treatment as shown in the precious section can be applied for the frame without an effective field, whenever $\omega^{\text{eff}}_{x} \not \approx 0 \not \approx \omega^{\text{eff}}_{y}$.
Because every spin has its unique interaction frame, we align the effective field vectors $\vec{\omega}^{(i)}_{\textrm{eff}}$ with the z-axis for each spin $(i)$ with an rotation $^iR_{\mu\nu}$
As a result the effective field vectors will be all aligned with the z-axis and hence we can apply the treatment as described in the main text and obtain the (rotated) scaling factor $\tilde{C}^{(ij)}_{\mu\nu}$.
As before we account for the mixing of $\widehat{A}_{\mu \nu}(\Omega)$ due to the rotations by inverting them at the end as  
\begin{align}
  \label{eq:inverse_rotation_eff}
  C^{(ij)}_{\mu\nu}=\sum_{\chi\chi'} {^iR^{(-1)}_{\chi\mu}}\, {^jR^{(-1)}_{\chi'\nu}} \tilde{C}^{(ij)}_{\chi\chi'}
\end{align}